# HELIUM BEHAVIOR IN OXIDE NUCLEAR FUELS: FIRST PRINCIPLES MODELING

D. Gryaznov[1], S. Rashkeev[2], E.A. Kotomin[3], E. Heifets[1], Y. Zhukovskii[3]

[1]European Commission, Joint Research Center, Institute for Transuranium Elements, 76125 Karlsruhe, Germany

[2]Center for Advanced Modeling & Simulations, Idaho National Laboratory, Idaho Falls, ID 83415, USA

[3]Institute of Solid State Physics, University of Latvia, Kengaraga 8, Riga LV-1063, Latvia

Corresponding author: Dr. Denis Gryaznov, e-mail: gryaznov@mail.com, Current address: Max Planck Institute for Solid State Research, Heisenbergstr. 1, Stuttgart, D-70569

**Abstract.** $UO_2$ and $(U,Pu)O_2$ solid solutions (the so-called MOX) nowadays are used as commercial nuclear fuels in many countries. One of the safety issues during the storage of these fuels is related to their self-irradiation that produces and accumulates point defects and helium therein.

We present density functional theory (DFT) calculations for $UO_2$, $PuO_2$ and MOX containing He atoms in octahedral interstitial positions. In particular, we calculated basic MOX properties and He incorporation energies as functions of Pu concentration within the spin-polarized, generalized gradient approximation (GGA) DFT calculations. We also included the on-site electron correlation corrections using the Hubbard model (in the framework of the so-called DFT+U approach). We found that $PuO_2$ remains semiconducting with He in the octahedral position while $UO_2$ requires a specific lattice distortion. Both materials reveal a positive energy for He incorporation, which, therefore, is an exothermic process. The He incorporation energy increases with the Pu concentration in the MOX fuel.

## Introduction

The MOX solid solutions are of considerable interest for the nuclear engineering due to the issues of the plutonium re-use, separation and/or storage [1].

$PuO_2$ and $UO_2$ are known to have the fluorite structure within a broad temperature range. Although their lattice constants differ by ~1.3 % only, their magnetic properties at low temperatures are different. $UO_2$ obeys a 3-k magnetic structure and oxygen sublattice distortion [2] at temperatures below the Neel temperature of 30.8$^o$K [3]. The magnetic moment on U ion is 1.75 $\mu_B$ [4] which is close to that anticipated from the 5f-orbital occupation. In $PuO_2$, however, the magnetic susceptibility measurements do not give any substantial magnetic moment on Pu ion and suggest its temperature independence [5]. The magnetic properties of $PuO_2$ or MOX are out of the scope of the present study – we analyze the basic bulk properties only. It is, for example, known that the MOX lattice constant obeys Vegard's law [6] as a function of Pu concentration, and the thermal conductivity is smaller than in $UO_2$ [7]. Both $UO_2$ and $PuO_2$ are so-called Mott insulators with the band gap of 2.0 eV [8] and 1.8 eV [9], respectively.

$UO_2$ was intensively studied (including bulk properties and defects behaviour) by the DFT approaches [10-15]. Some theoretical efforts based on the DFT+U approach [16-17] and the hybrid exchange-correlation functional [18] were also recently made to study the electronic structure of $PuO_2$. In particular, the anti-ferromagnetic (AFM) structure in $PuO_2$ was suggested to be energetically more favourable than the ferro-magnetic (FM) one. In these studies the calculated band gap was close to the experimental value. Two different modifications of the DFT+U method were applied in [16-17] which is the issue carefully analyzed in [19].

He is accumulated in MOX mostly as a result of actinide alpha- decay which seriously affects fuel mechanical and thermal properties. In this paper, we study He incorporation into MOX for the first time by using ab initio calculations.

## Computational details

In present simulations we employed the VASP 4.6 code [20-21]. The calculations are based on the projector augmented wave (PAW) method combined with the scalar relativistic pseudopotentials [22]. The core states (78 electrons for Pu, 76 electrons for U and 2 electrons for O) were included into the pseudopotentials. The Hubbard model was used to treat strong correlations within the so-called DFT+U method, and the relevant Hamiltonians as available in VASP were applied. The simplified rotationally invariant Dudarev's form for Hubbard correction, which deals exclusively with the difference between an on-site Coulomb U and exchange J parameters leading to $U_{eff}$ = U -J, was used to study $UO_2$ [11]. In contrast to $UO_2$, $PuO_2$ shows a significant role of the exchange part requiring the Liechtenstein form [23] of the Hubbard correction. In general, $UO_2$ is less sensitive to the choice of the functional, and we also used the DFT+U approach within the Liechtenstein form for the MOX with the U, J parameters separately found for $UO_2$ and $PuO_2$. Namely, the parameters U of 4.6 and 3.0 eV and the exchange parameters of 0.5 and 1.5 eV were used for $UO_2$ and $PuO_2$, respectively. The parameters used for $UO_2$ are in agreement with the experimental study of Baer [8] and previous *ab initio* calculations [11]. The parameters for $PuO_2$ were carefully determined by calculating a number of properties and varying them in a wide range [19, 24].

The calculations employed a 1-k AFM structures for $UO_2$ and $PuO_2$ on the basis of experimental data [5] and previous calculations [16-17]. Such a magnetic structure means that spins are pointed in the [001] direction in the FM planes which leads to the tetragonal distortion of the cubic oxide lattice. The AFM structure for $PuO_2$ was also confirmed to be lower in energy than the FM one. The same [001] magnetic structure was also applied to the solid solution. The comparisons were also made with the FM structure for different Pu concentrations. The relevant unit cells were fully relaxed, including both the lattice parameters and magnetic moments.

The calculations of He atom in octahedral position involved two types of supercells, namely the 13 atom and 25 atom supercell. Such supercells are constructed from the conventional fluorite unit cell. Also, these supercells are particularly useful for the MOX calculations allowing us to treat the following concentrations of Pu: 12.5% (MOX 12.5), 25% (MOX 25), 50% (MOX 50), 75% (MOX 75), 87.5% MOX (87.5). However, we observed some instabilities of the DFT+U method for the MOX calculations. It is reflected in the scattering behaviour of calculated bulk properties, like magnetic moment, as functions of Pu concentration and can easily bring the system to a wrong conducting state [19]. This issue was already discussed in the literature [17]. To avoid this problem, we lifted the symmetry constraints for the MOX calculations (with and without He), which significantly improves the dependence of the calculated properties on the Pu concentration.

The plane wave cut-off energy was fixed at 520 eV except for supercells with He where it was increased up to 621 eV. The exchange-correlation functional as applied within the DFT+U scheme is the one of Perdew, Becke, and Ernzerhof (PBE) [25]. The Brillouin zone integration employed 10x10x8 Monkhorst-Pack k-point mesh [26] for the tetragonal unit cell of $PuO_2$, 6x6x6 for the conventional unit cell of MOX and 8x8x8 for the conventional unit cell of $PuO_2$ whereas the electron occupancies were determined with the Gaussian method using smearing parameter of 0.1 eV for Pu containing oxides. The details of the calculation accuracies for pure $UO_2$ calculations are provided elsewhere [27]. The density of states, however, was re-calculated with the tetrahedron metod as suggested by Bloechl [28].

**Results and discussion**

1. **Calculations on $UO_2$ and $PuO_2$ bulk**

As a result of the lattice relaxation, the optimized lattice constant *c* (table 1) along the z-direction in $UO_2$ with alternating U spins is reduced by ~1% as compared to the two other directions a(b). The bulk modulus (B) was estimated by fitting parameters of the Birch-Murnaghan equation of state [29] to the adiabatic energy profiles (the total energy per unit cell *vs.* unit cell volume) calculated with the VASP code. The cohesive energy (bulk modulus) $E_{coh}$ (B) only slightly exceeds (is smaller in comparison to) the experimental value, which is in agreement with the general trend for the GGA functionals. The effective U ion charges found by using the topological (Bader) method [30] considerably differ from the formal charges ($U^{4+}$, $O^{2-}$), thus, suggesting a partly covalent nature of $UO_2$ bonding. The magnetic moments $\mu$ on the U ion confirm this picture. However, it is slightly larger than the experimental value of 1.74 $\mu_B$. Notice that the results in table 1 do not include the spin orbit coupling effects. The DFT+U approach gives the $UO_2$ band gap $E_g \sim 2$ eV, consistent with the experiment (fig. 1a). As it is expected, the U 5f electrons form a separate (localized) band below the Fermi energy with a small contribution of O 2p orbitals, which leads to the U–O partial bond covalence. This was observed also in the local combination of atomic orbitals (LCAO) computations with the Gaussian code [18] performed using the hybrid functional.

On the basis of the procedure described in [27], we observed that the lowest energy structure within the [001] magnetism for $UO_2$ is of orthorhombic symmetry (table 1). This result is very much in line with experimental observations on the distortion mechanisms in $UO_2$. The orthorhombic structure itself is characterized by comparable bulk properties together with the tetragonal one; however, it appeared to be important for the He studies (see discussion below).

A similar comparison of experimental and theoretical results for basic bulk properties is also possible for $PuO_2$. As mentioned above, the hybrid functionals and DFT+U method applied clearly indicate an insulating character of $PuO_2$. The AFM structure was also confirmed which is obviously inconsistent with its measured magnetic properties. We will report elsewhere on the role of spin orbit effects [19] to explain this discrepancy. The DFT+U scheme relying on the Liechtenstein functional was used in [17] whereas the Dudarev functional was applied in [16]. Despite using the same values for the U-parameter, they produce slightly different values for $E_g$ and B but predict a comparable contribution of Pu 5f- and O 2p-electrons to the density of states (DOS). Nevertheless, this is not fully confirmed experimentally despite $E_g$ being close to the experimental value. Also, the hybrid functional as applied in [18] demonstrates the same picture with $E_g$, however, overestimated.

Our calculations for $PuO_2$ by means of the DFT+U method demonstrate that the role of exchange parameter may not be neglected, and it is better to properly tune this parameter [19]. We have found that with U = 3.0 and J = 1.5 eV a very good agreement with the experimental results is obtained. In table 1 we present the newly calculated values of basic bulk properties of $PuO_2$. The AFM structure reveals the tetragonal symmetry with two lattice parameters a(b), c. The equilibrium unit cell volume is very close to that found by Jomard [17]. The cohesive energy $E_{coh}$ is similar to that in $UO_2$. The effective charge is far from the formal charge indicating a contribution of covalency to the bonding like in $UO_2$. The high magnetic moment is close to the value suggeted by Jomard [17] whereas the band gap is slightly underestimated in comparison with the experimental value. Interestingly, the bulk modulus B and band gap $E_g$ are underestimated in comparison to both the esperimental values and previous calculations. The Pu 5f-electrons lie at the Fermi energy (fig. 1b). The position and contribution of 5f-electrons is changed, leading to their clear localized behavior, in comparison to previous theoretical works.

## 2. Calculations on MOX bulk

As mentioned above, all the DFT+U MOX calculations with symmetry constraints relaxed to wrong conducting ground states for all Pu concentrations. We, therefore, decided to lift the symmetry constraints and allowed the system to converge to the energetically most stable state. Such an approach leads to insulating solid solutions for all Pu concentrations.

The MOX calculations were done with the same U, J-parameters as for pure $PuO_2$. For each of the Pu concentrations we calculated the AFM, FM and ferrimagnetic (FEM) solutions, in order to find the minimum energy structure. Such solutions represent different total magnetic spin moments of the system: the total magnetic spin moment of 1, -1 $\mu_B$ is possible for MOX 12.5, MOX 25, MOX 75, MOX 87.5 per respective unit cell (i.e. 24 atoms unit cell in the case of 12.5% (87.5%) concentration and 12 atoms unit cell in the case of 25% (75%) concentration). The zero total magnetic spin moment requires using a 24 atom supercell for MOX 25 (MOX 75) and 96 atom supercell for MOX 12.5 (MOX 87.5).

The energy difference between the FM and AFM (FEM) states vary with the Pu concentration. We observed that the FEM structure is lower in energy for small Pu concentrations whereas the concentrations higher than 50% are characterized by the FM structure. In fig. 2 the volume of the unit cell is plotted as a function of Pu content. The calculated results agree with the experimental result [6]. Both the theoretical estimate and experimental considerations agree on Vegard's law validity [35] for MOX with the volume decreasing with the Pu content. The theoretical volume is overestimated in comparison to the experimenal one, what can have different grounds. First, the DFT calculation itself typically overestimates the volumes of materials. Second, the symmetry of calculated unit cells lowers and is not characterized by the single lattice parameter like in experiments.

## 3. He incorporation into $UO_2$, $PuO_2$ and MOX

As observed earlier [27], $UO_2$ reveals the conducting state when He occupies the high symmetry octahedral position in a cubic crystal. A solution to this problem can be found through the lifting the symmetry constraints (for details, see [27]). The obtained crystal structure has a *monoclinic* symmetry with the He position showing a reduced $C_{2v}$ point symmetry. This structure can be described as the tetragonal supercells sheared along the base diagonal. Still, a deviation from a cubic structure appears to be very small. Thus, the calculated lattice vectors for the 13 atom supercell in the basal plane are $a = b = 5.564$ Å (with the angle between them $\gamma = 90.08°$) whereas $c = 5.563$ Å (with the angle with each of the basal lattice vectors is $\alpha = \beta = 90.17°$).

The DOS for a 13 atom *monoclinic* supercell is shown in Fig. 3a. The total DOS for the same supercell of a pure orthorhombic $UO_2$ (without He atom) is also added there for a comparison. The band gap in the supercells with He is ~2.5 eV, which is ~0.5 eV higher than that for pure $UO_2$. He *1s*-states lie ~7 eV below O *2p*-states, which is much lower than valence electrons. Like in a pure $UO_2$, the U *5f* electrons contribute to both the highest valence band and the lowest conduction band. The introduction of a He atom into an interstitial position and the resulting monoclinic distortion leads to the higher-energy shift by ~0.5 eV of both the O *2p* valence band and the conduction band.

To estimate the He incorporation energy, we had to use supercells for a perfect $UO_2$ crystal with the same reduced (orthorhombic) symmetry. The calculated incorporation energies turn out to be now 0.59 eV in a 13 atoms supercell (table 2). This value is smaller than that found in the local density approximation studies by Crocombette [10] for a metallic $UO_2$. As seen from [27] there is still a dependence on the supercell size and these values may also be corrected to include an important van der Waals contribution which is, however, out of the scope of the present study.

Introduction of He into the octahedral position of $PuO_2$ does not lead to the instability and the conducting state observed for $UO_2$. $PuO_2$ with He is characterized by a slightly reduced band gap (fig. 3b). The position of He 1s-electrons is similar to that in $UO_2$. The Pu 5f-electrons are also the outermost electrons and the position of other bands does not change in comparison to the defect-free bulk material. The incorporation energy for He in $PuO_2$ and MOX is also given in table 2. It confirms that the incorporation energy may increase with the Pu content. All the calculations on He in MOX fuel with the Pu concentrations below 25 % indicate an insulating behavior with a band gap slightly reduced (~ 1 eV) in a comparison with both pure $UO_2$ and $PuO_2$ (fig. 4). The 5f-electrons remain relatively localized for small Pu concentrations. However, they mostly comprise electrons of U whereas Pu contributes to the conduction band.

## Conclusions and perspectives

1) The host crystalline symmetry reduction is particularly important for calculations of He incorporation energy into octahedral interstitial position in actinide dioxides. In particular, in $UO_2$ properly calculated He incorporation energy turns out to be reduced in a comparison to high symmetry with conducting states.
2) $PuO_2$ reveals considerably larger He incorporation energy than $UO_2$.
3) The calculations on MOX solid solutions require proper analysis of the ground state. Lifting the symmetry constraints resolves the problem but indicates complex lattice distortion.
4) The calculations discussed here were performed for He in the octahedral position only. Our future studies will include larger supercells and points defects on both the U and Pu MOX sublattices, in order to analyze the role of defects in the He behavior.

**Acknowledgments**

This study was supported by the ISSP-ITU service contract No 211424-08-2008 F1ED KAR LV), and the supercomputers of the EMS Laboratory at PNNL (Project Nr. 25592). Authors are greatly indebted to R. Caciuffo, Th. Gouder, P. van Uffelen, R. A. Evarestov, R. Konings for very valuable discussions. D.G. acknowledges also the EC for support in the frame of the Program "Training and mobility of researchers".

**Literature**

[1] Plutonium Management in the Medium Term: A review by the OECD/NEA working party on the physics of plutonium fuels and innovative fuel cycles (WPPR), 2003, ISBN 92-64-02151-5

[2] R. Caciuffo, G. Amoretti, P. Santini, G. H. Lander, J. Kulda, P. de V. Du Plessis, Phys. Rev. B 59 (1999) 13892

[3] B. C. Frazer, G. Shirane, D. E. Cox, Phys. Rev. 140 (4A) (1965) A1448

[4] J. Faber, Jr, G. H. Lander, Phys. Rev. B 14 (3) (1976) 1151

[5] M. Colarieti-Tosti, O. Eriksson, L. Nordström, J. Wills, M. S. S. Brooks, Phys. Rev. B 65 (2002) 195102

[6] P. Martin, S. Grandjean, C. Valot, G. Carlot, M. Ripert, P. Blanc, C. Hennig, J. Alloys Comp. 444-445 (2007) 410

[7] J. K. Fink, J. Nucl. Mater. 279 (2000) 1

[8] Y. Baer, J. Schoenes, Solid State Commun. 33(1980) 885

[9] C. E. McNeilly, J. Nucl. Mater. 11 (1) (1964) 53

[10] J.-P. Crocombette, J. Nucl. Mater. 305 (2002) 29

[11] S. L. Dudarev, G. A. Botton, S. Y. Savrasov, Z. Szotek, W. M. Temmerman, A. P. Sutton, Phys. Status Solidi A 166 (1998) 429

[12] Y. Yun, O. Eriksson, P.M. Oppeneer J. Nucl. Mater. 385 (2009) 509

[13] H. Y. Geng, Y. Chen, Y. Kaneda, M. Kinoshita Phys. Rev. B 75 (2007) 054111

[14] M. Iwasawa, Y. Chen, Y. Kaneta, T. Ohnuma, H.-Y. Geng, M. Kinoshita Mater. Trans. 47 (11) (2006) 2651

[15] F. Gupta, G. Brillant, A. Pasturel Phil. Mag. 87 (2007) 2561

[16] B. Sun, P. Zhang, X.-G. Zhao J. Chem. Phys. 128 (2008) 084705

[17] G. Jomard, B. Amadon, F. Bottin, M .Torrent Phys. Rev. B 78 (2008) 075125

[18] I. D. Prodan, G. E. Scuseria, R. L. Martin Phys. Rev. B 73 (2006) 045104

[19] D. Gryaznov, D. Sedmidubsky (in preparation)

[20] G. Kresse, J. Furthmueller *VASP the Guide*, University of Vienna, 2007

[21] G. Kresse, J. Furthmüller Comp. Mater. Sci. 6 (1996) 15

[22] G. Kresse, D. Joubert Phys. Rev. 59 (1999) 1758

[23] A. I. Liechtenstein, V. I. Anisimov, J. Zaane Phys. Rev. B 52 (1995) R5467

[24] D. Gryaznov, D. Sedmidubsky, E. Heifets J. Phys.: Conf. Ser. (accepted)

[25] J. P. Perdew, K. Burke, M. Ernzerhof Phys. Rev. Lett. 77 (1996) 3865

[26] H. J. Monkhorst, J. D. Pack Phys. Rev. B 13 (1976) 5188

[27] D. Gryaznov, E. Heifets, E. Kotomin Phys. Chem. Chem. Phys. 11 (2009) 7241

[28] P. E. Bloechl, O. Jepsen, O. K. Andersen Phys. Rev. B 49 (1994) 16223

[29] Birch Phys. Rev. 71 (1947) 809

[30] R. Bader, *Atoms in Molecules: A Quantum Theory*, Oxford University Press, New York, 1990

[31] M. Idiri, T. Le Behan, S. Heathman, J. Rebizant Phys. Rev. B 70 (2004) 014113

[32] R. Caciuffo, N. Magnani, P. Santini, S. Carretta, G. Amoretti, E. Blackburn, M. Enderle, P. J. Brown, G. H. Lander J. Mag. Mater. 310 (2007) 1698

[33] W. H. Zachariasen Acta Cryst. 2 (1949) 388

[34] G. Raphaell, R. Lallement Solid State Commun. 6 (1968) 383

[35] A. R. Denton, N. W. Ashcroft Phys. Rev. A 43 (1991) 3161

TABLE 1. A comparison of $UO_2$ and $PuO_2$ bulk properties (lattice constants (*a, b, c*), cohesive energy $E_{coh}$, bulk modulus B, U effective (Bader) charges $Q_{eff}(U)$ and magnetic moments μ(U), and band gap $E_g$) obtained with the PBE exchange-correlation functional $E_{xc}$ (see the text for details). X = U or Pu , respectively.

| | $UO_2$ | | | | $PuO_2$ | | |
|---|---|---|---|---|---|---|---|
| | Tetragonal phase | Orthorhombic phase | Expt | Other calculations | Tetragonal phase | Expt | Other calculations |
| a (b) Å | 5.567 | 5.576 (5.575) | 5.47[c] | 5.55[f], 5.37[a], 5.45[b], 5.46[b] | 5.402 | 5.386[h] | 5.47[l] 5.396[m] |
| c, Å | 5.512 | 5.508 | | 5.47[f] | 5.513 | - | - |
| $E_{coh}$, eV | 23.0 | 23.0 | 22.3[a] | 22.2[a] | 21.7 | - | - |
| B, GPa | 180 | - | 207[e] | 173[a], 219[b] | 151 | 178[e] | 199[g] 220[m] 184[l] |
| $Q_{eff}(X)$, e | 2.6 | 2.7 | | | 2.4 | - | - |
| μ(X), $\mu_B$ | 2.0 | 2.0 | 1.74[c] | ~1.9[a] | ±3.8 | Non-magnetic[j] | 3.89[g] |
| $E_g$, eV | 1.94 | 2.0 | 2.0[d] | 1.3[a], 3.13[b], 2.64[b] | 1.5 | 1.8[i] | 2.2[g] 2.64[m] |

[a)] Ref. 11, [b)] Ref. 32, [c)] Ref. 2, [d)] Ref. 8, [e)] Ref. 31, [f)] Ref. 14

[g)] Ref. 17, [h)] Ref. 33, [i)] Ref. 9, [j)] Ref. 34, [k)] Ref. 32, [l)] Ref. 16, [m)] Ref. 18

TABLE 2. He incorporation energies for $UO_2$, $PuO_2$ and MOX fuels for selected concentrations of Pu. The energy was calculated for the distorted structure of $UO_2$ and lifting the symmetry constraints of Pu contained fuels (see the text for details). The energies as given for $UO_2$ and $PuO_2$ correspond to the 13 atoms supercells. The energies are given for the energetically stable configuration of MOX fuel.

| Fuel | $UO_2$ | $PuO_2$ | MOX 12.5 | MOX 25 |
|---|---|---|---|---|
| Incorporation energy, eV | 0.59 | 1.02 | 0.51 | 1.06 |

## Figure captions:

FIG. 1. The total and partial DOS calculated by using the PBE functional for tetragonal a) $UO_2$ and b) $PuO_2$. The Fermi energy is taken as zero. The negative values correspond to spin-down electrons.

FIG. 2. The relaxed unit cell volume for MOX as a function of Pu concentration.

FIG. 3. The DOS calculated for a) $UO_2$ and b) $PuO_2$ with He atom incorporated into the octahedral position of the 13 atom supercell. The distorted structures are shown for $UO_2$ with and without He, whereas the partial DOS is provided for $PuO_2$ with incorporated He.

FIG. 4. The DOS calculated for MOX 12.5 and MOX 25 and He atom in the octahedral position.

**a)**

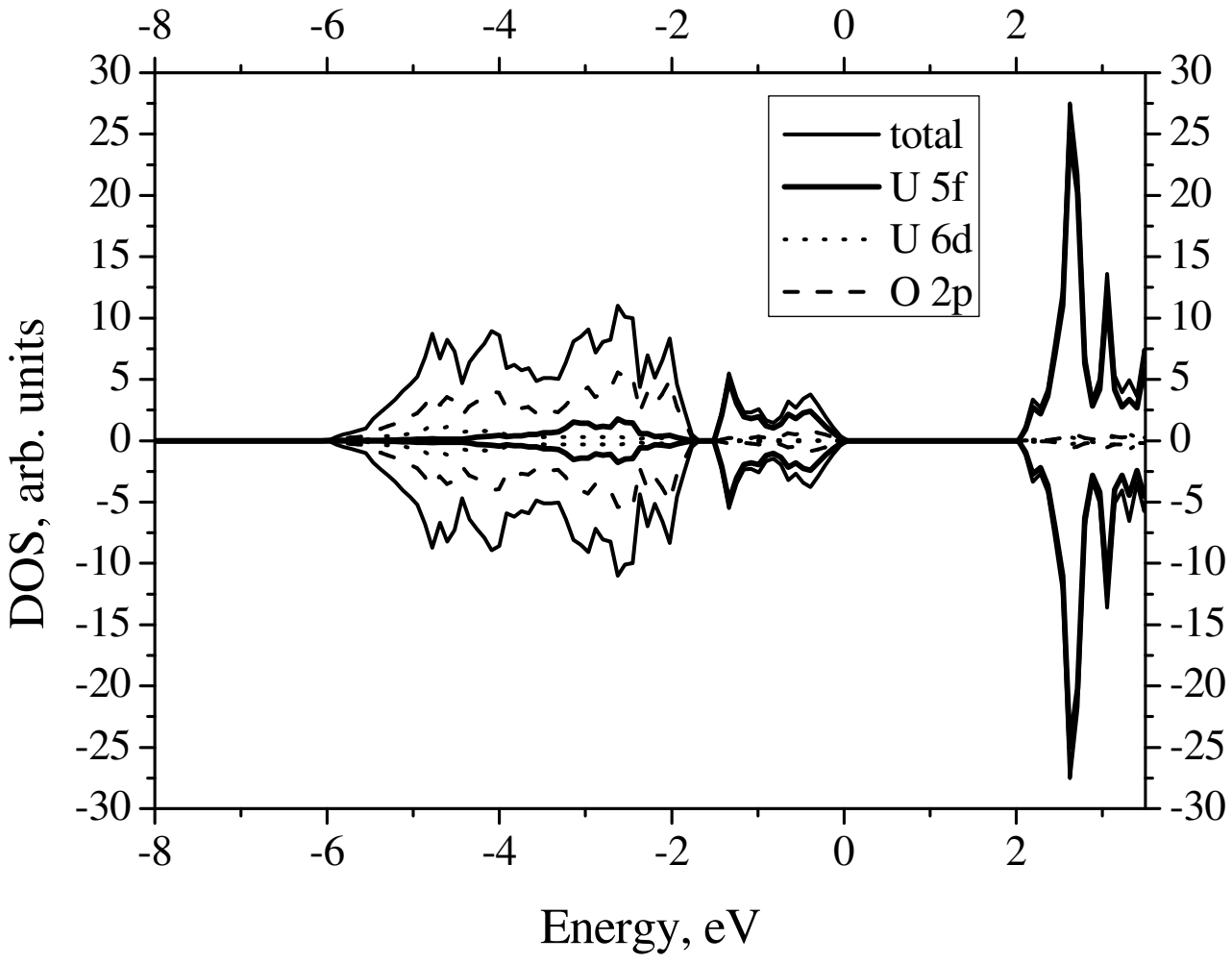

**b)**

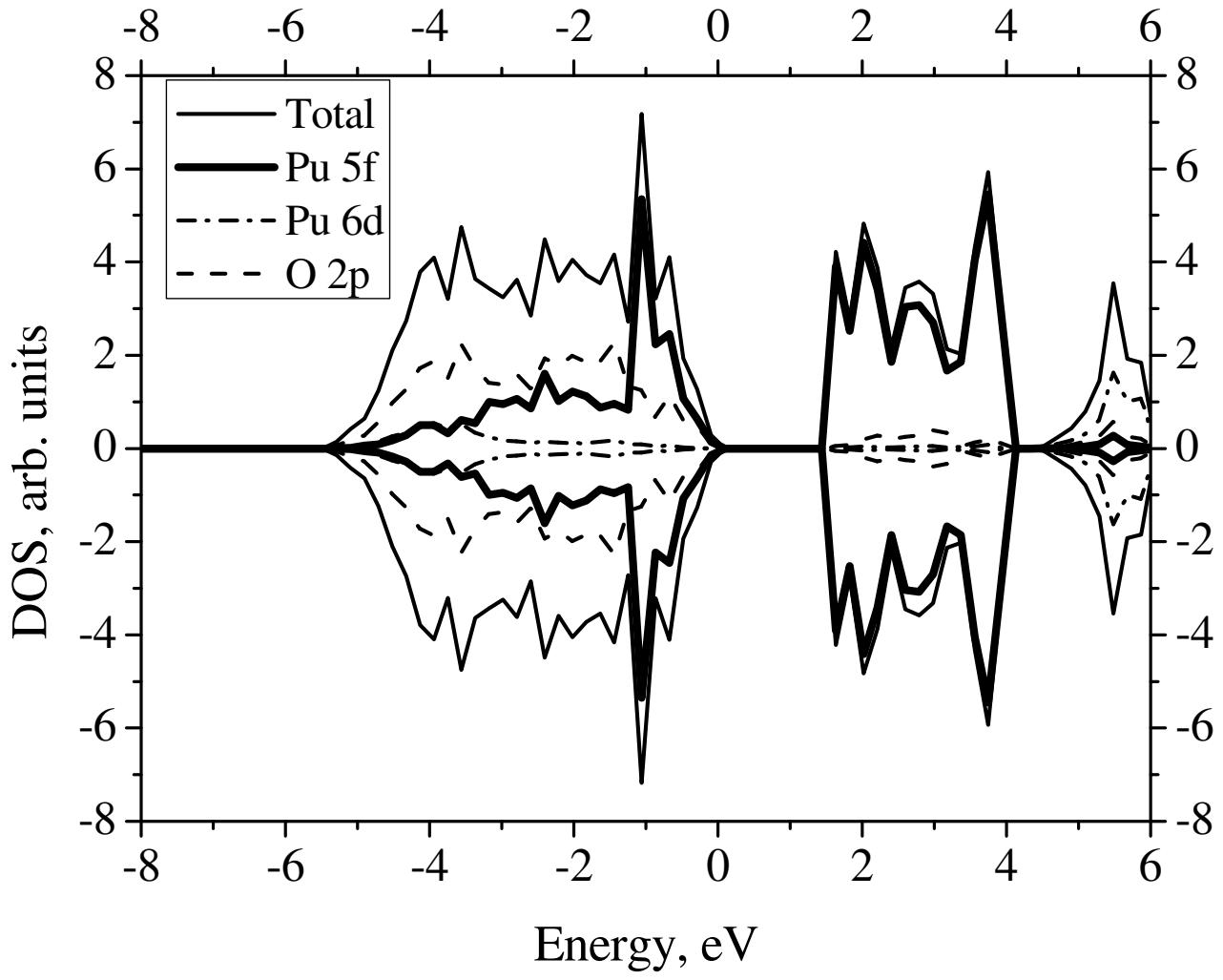

FIG. 1

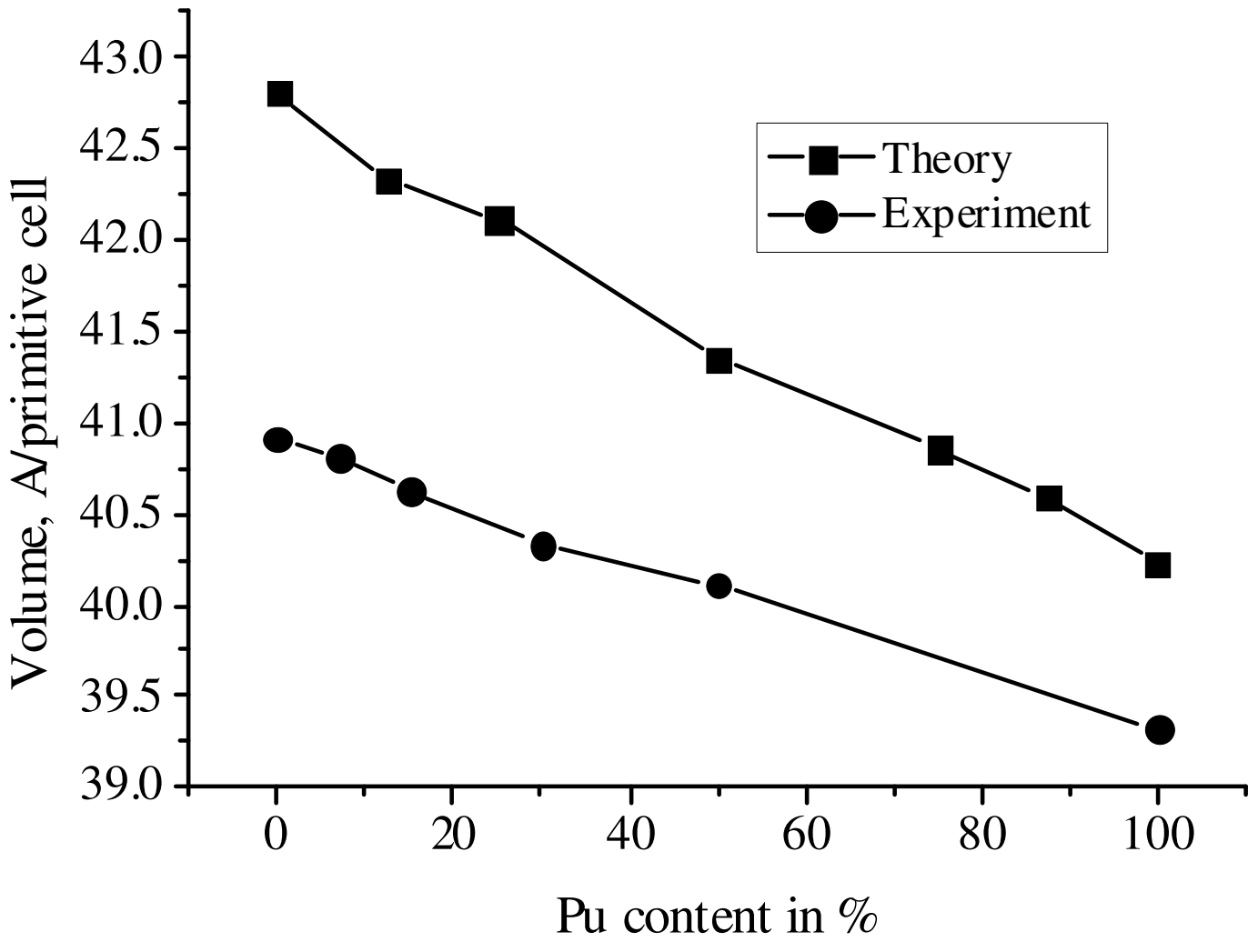

FIG. 2

a)

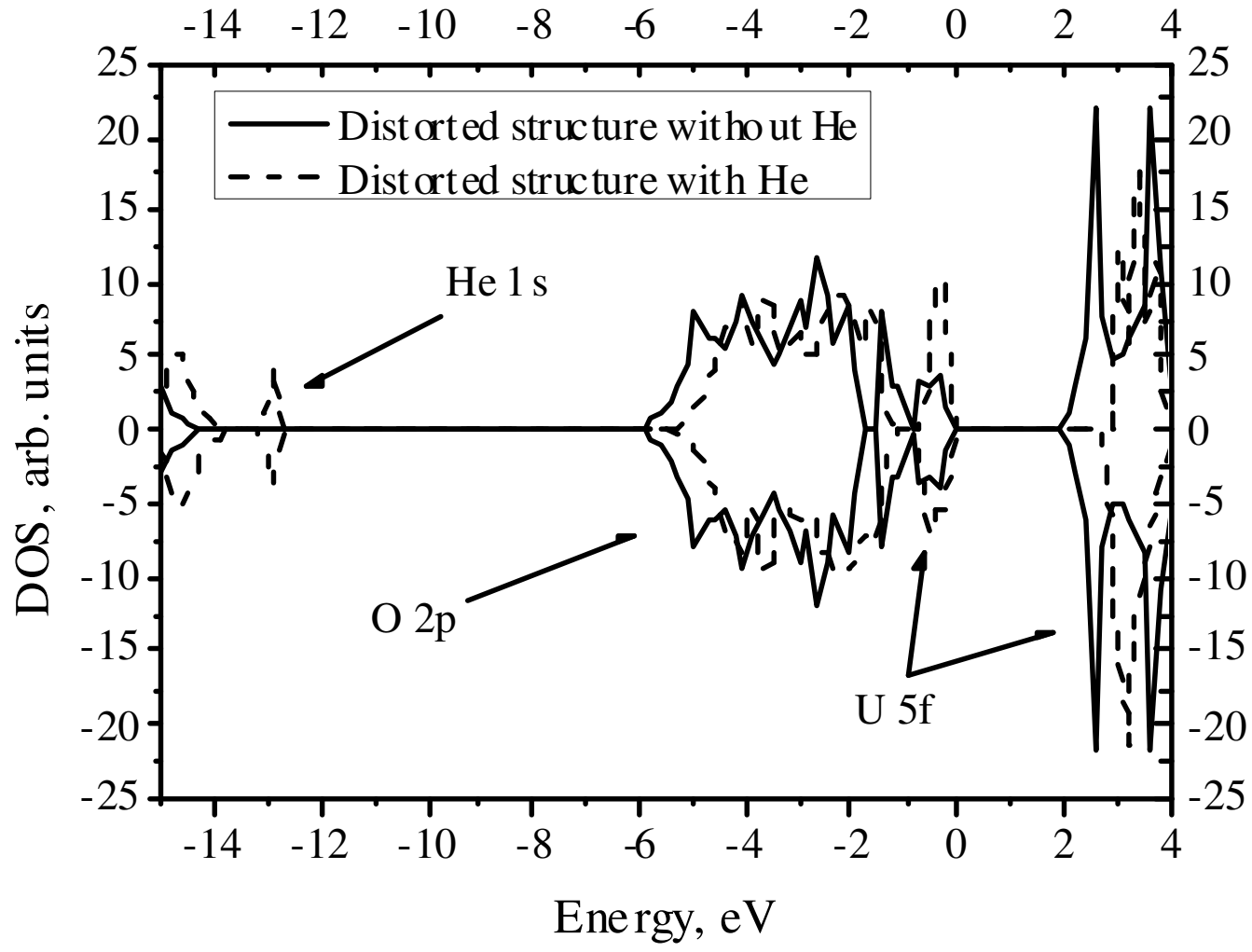

Distorted structure without He
Distorted structure with He
He 1s
O 2p
U 5f
DOS, arb. units
Energy, eV

**b)**

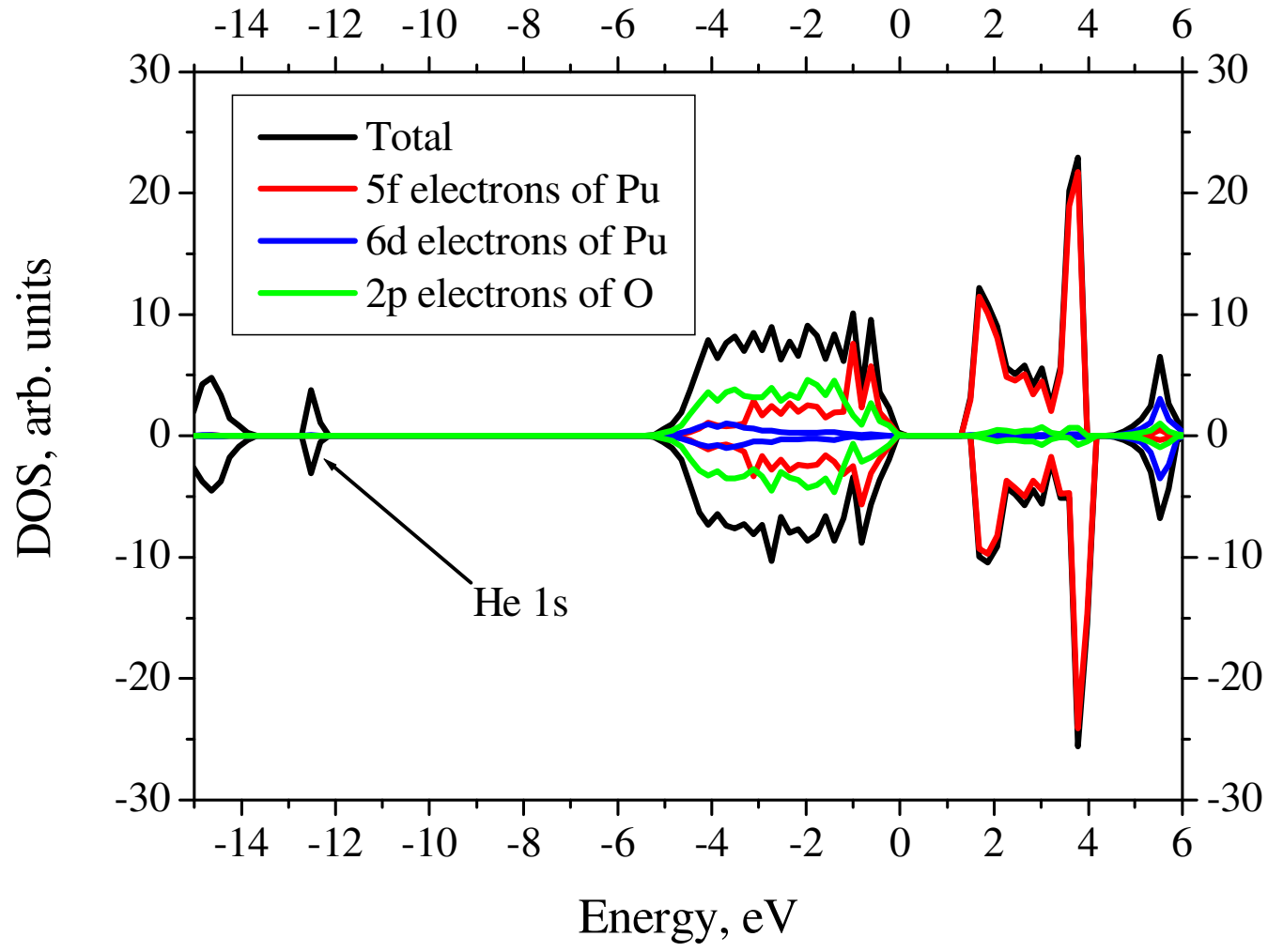

Total
5f electrons of Pu
6d electrons of Pu
2p electrons of O
He 1s
DOS, arb. units
Energy, eV

FIG. 3

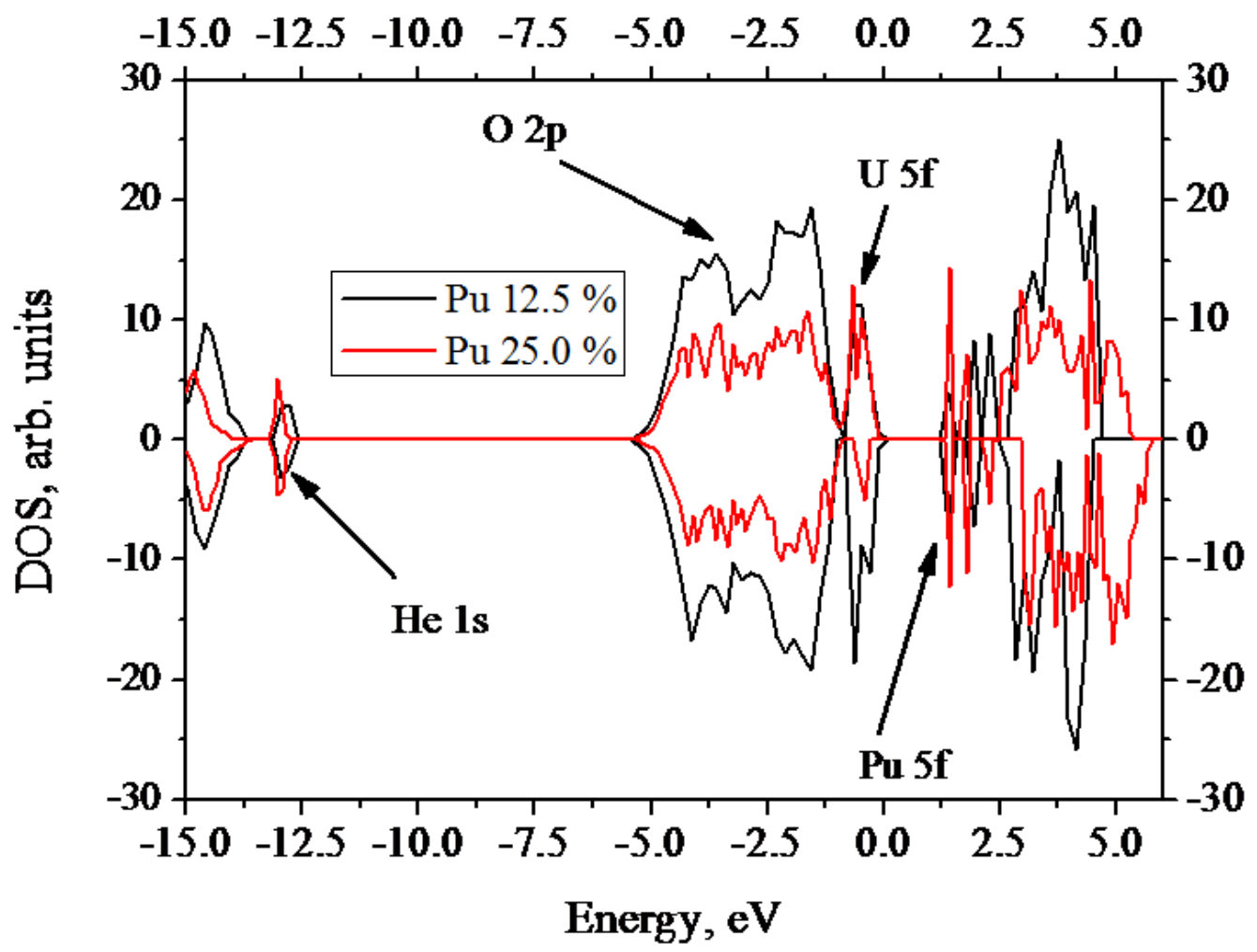

-15.0 -12.5 -10.0 -7.5 -5.0 -2.5 0.0 2.5 5.0
30
20
10
0
-10
-20
-30
O 2p
U 5f
Pu 12.5 %
Pu 25.0 %
He 1s
Pu 5f
DOS, arb. units
Energy, eV

FIG. 4